\documentclass[nofootinbib,prd,preprintnumbers,superscriptaddress,aps]{revtex4}
\usepackage[utf8]{inputenc}
\pdfoutput=1

\usepackage{comment} 
\usepackage{graphicx}
\usepackage{epsfig}
\usepackage{bm}
\usepackage{amssymb}
\usepackage{float}
\usepackage{amsmath}
\usepackage{dcolumn}
\usepackage{cancel}
\usepackage[colorlinks]{hyperref}
\usepackage[usenames, dvipsnames]{color}

\hypersetup{
     breaklinks=true, 
    pdfstartview={FitH},  
    colorlinks=true, 
    linkcolor=blue,  
    citecolor=red,  
    filecolor=magenta,  
    urlcolor=blue, 
    anchorcolor=green,  
    linktocpage=true
}

\providecommand{\U}[1]{\protect\rule{.1in}{.1in}}

\newcommand{\be}{\begin{equation}}
\newcommand{\ee}{\end{equation}}

\newcommand{\mincir}{\raise
-3.truept\hbox{\rlap{\hbox{$\sim$}}\raise4.truept\hbox{$<$}\ }}
\newcommand{\magcir}{\raise
-3.truept\hbox{\rlap{\hbox{$\sim$}}\raise4.truept\hbox{$>$}\ }}

\ifx\pdfoutput\relax\let\pdfoutput=\undefined\fi
\newcount\msipdfoutput
\ifx\pdfoutput\undefined\else
\ifcase\pdfoutput\else
\ifx\paperwidth\undefined\else
\ifdim\paperheight=0pt\relax\else\pdfpageheight\paperheight\fi
\ifdim\paperwidth=0pt\relax\else\pdfpagewidth\paperwidth\fi
\fi\fi\fi

\hypersetup{colorlinks=true,
	breaklinks=true,
	pdfstartview=Fit,
	linkcolor=blue,
	citecolor=blue,
	urlcolor=blue}

\begin{document}
\title{Modified cosmology through generalized mass-to-horizon entropy: implications for structure growth and primordial gravitational waves}

\author{Giuseppe Gaetano Luciano}
\email{giuseppegaetano.luciano@udl.cat}
\affiliation{Departamento de Qu\'{\i}mica, F\'{\i}sica y Ciencias Ambientales y del Suelo, Escuela Polit\'ecnica Superior -- Lleida, Universidad de Lleida, Av. Jaume II, 69, 25001 Lleida, Spain}

\begin{abstract} 
In the framework of entropic cosmology, entropic forces arising at the cosmological horizon have been proposed as an alternative mechanism to explain the Universe’s current accelerated phase. However, recent studies have shown that, under the Clausius relation and assuming a linear mass-to-horizon (MHR) relation, all entropic force models reduce to the original Bekenstein-Hawking formulation, regardless of the specific form of the horizon entropy. As a result, they inherit the same observational limitations in accounting for cosmic dynamics. To address this issue, a generalized MHR has been introduced, 
providing the foundation for a modified cosmological scenario rooted in the  gravity-thermodynamics conjecture. In this work, we explore the implications of this generalized framework for early-Universe dynamics. Specifically, we analyze the growth of matter perturbations within the spherical Top-Hat formalism in the linear regime, showing that the density contrast profile is significantly influenced by the modified background dynamics predicted by the model.
Moreover, considering the sensitivity of upcoming gravitational wave detectors in the sub-$10^3\,\mathrm{Hz}$ range, we examine the impact on the relic abundance of Primordial Gravitational Waves (PGWs), identifying parameter regions where deviations from standard cosmology may arise through an enhanced PGW spectrum.
\end{abstract}

\maketitle

\section{Introduction}
\label{Intro}
Recent astrophysical observations, including supernova luminosity distances \cite{SupernovaSearchTeam:1998fmf,SupernovaCosmologyProject:1998vns}, cosmic microwave background anisotropies \cite{COBE:1992syq,WMAP:2003ivt} and large-scale structure surveys \cite{2DFGRS:2001zay,SDSS:2003eyi,BOSS:2016wmc}, provide compelling evidence that the Universe has undergone two distinct epochs of accelerated expansion: an early inflationary phase and the late-time acceleration. Addressing these phenomena has led to the development of two primary theoretical strategies.

The first strategy involves modifying the geometric framework of gravity itself. Instead of strictly adhering to Einstein’s original formulation, extensions of the Einstein–Hilbert action are considered, leading to a broad class of models collectively known as modified gravity theories~\cite{Capozziello:2011et}. On the other hand, a conceptually distinct approach retains general relativity as the governing theory of gravity but modifies the matter sector. In this context, the introduction of new dynamical components such as scalar fields (e.g., the inflaton) or dark energy fluids plays a central role in driving cosmic acceleration~\cite{Olive:1989nu,Bartolo:2004if,Copeland:2006wr,Cai:2009zp,CosmoVerse:2025txj}.

An alternative and increasingly influential perspective posits a deep connection between gravitational dynamics and thermodynamics~\cite{Jacobson:1995ab,Padmanabhan:2003gd,Padmanabhan:2009vy}. Within this framework, the Universe is modeled as a thermodynamic system bounded by the apparent horizon and the field equations can be recovered by applying the first law of thermodynamics to this boundary~\cite{Frolov:2002va,Cai:2005ra,Akbar:2006kj,Cai:2006rs}. This thermodynamic derivation remains valid not only for standard general relativity but also across a variety of modified gravity models, provided the corresponding entropy-area relations are properly generalized (see, e.g., \cite{Paranjape:2006ca,Akbar:2006er,Jamil:2009eb,Cai:2009ph}).

The notion of horizon entropy gains additional significance in the framework of entropic cosmology~\cite{Easson:2010av}, where thermodynamic arguments are employed to model the large-scale dynamics of the Universe. In this context, a concept closely tied to holographic entropy is that of \emph{entropic forces}, which emerge as effective contributions to the cosmological dynamics and are motivated by boundary terms in the Einstein–Hilbert action~\cite{Easson:2010av}. These additional terms are considered to account for the current accelerated expansion of the Universe. It is important to note that entropic cosmology 
is formulated within the framework of general relativity, where the Einstein field equations are applied to a Friedmann–Lema\^{\i}tre–Robertson–Walker (FLRW) background. This stands in contrast to Verlinde’s entropic gravity~\cite{Verlinde:2010hp}, in which gravity itself is interpreted as an emergent entropic phenomenon rather than a fundamental interaction.

Over the years, several generalized entropy measures have emerged as extensions of the semiclassical Bekenstein-Hawking entropy. These arise from non-standard statistical mechanics or from quantum and gravitational considerations on the holographic horizon. Notable examples include Rényi~\cite{renyi1961entropy}, Tsallis~\cite{Tsallis:1987eu,Tsallis:2009} and Sharma-Mittal~\cite{Sharma1975} entropies, which relax the assumption of extensivity; Kaniadakis entropy~\cite{kaniadakis2001non,Kaniadakis:2002zz,Luciano:2024bco}, rooted in relativistic statistical mechanics; and Barrow entropy~\cite{Barrow:2020tzx}, inspired by quantum-gravitational corrections to horizon geometry (see \cite{hanel2011comprehensive} for an axiomatic derivation of these generalized entropies and their corresponding distribution functions). 
These formulations recover the classical entropy in specific parameter limits. Accordingly, their integration into the thermodynamic description of gravity has sparked significant interest \cite{Lymperis:2018iuz,Saridakis:2020lrg,Nojiri:2019skr,Hernandez-Almada:2021rjs,Dheepika:2022sio,Jizba:2022icu,Lambiase:2023ryq,Jizba:2024klq,Ebrahimi:2024zrk,Nojiri:2025gkq}.

Nevertheless, a subtle but important issue has recently gained attention in the literature~\cite{Nojiri:2021czz,Gohar:2023lta}: whether it is theoretically consistent to generalize the entropy without simultaneously modifying other thermodynamic quantities. Some analyses argue that a change in entropy, by virtue of the first law, must be accompanied by corresponding adjustments to either the temperature or the internal energy of the system~\cite{Nojiri:2022sfd,Nojiri:2021czz}.

Another cosmology-driven viewpoint~\cite{Gohar:2023hnb,Gohar:2023lta} is inspired by the observation that as long as the Clausius relation is used to ensure thermodynamic consistency (i.e., to define the appropriate horizon temperature) and a linear mass-to-horizon relation (MHR) is assumed, any entropic force model becomes effectively indistinguishable from the original approach based on the Bekenstein entropy and Hawking temperature. This equivalence holds regardless of the specific entropy function adopted on the cosmological horizon. Consequently, all entropic cosmological models constructed under these assumptions inevitably inherit the same limitations as the Bekenstein–Hawking framework, notably its failure to accurately describe the observed cosmological dynamics at both the background and perturbative levels \cite{Basilakos:2012ra,Basilakos:2014tha}. To address this issue, a generalized MHR has been proposed, which in turn yields a modified expression for entropy that includes, as special cases, the Tsallis–Cirto \cite{Tsallis:2013}, Barrow and other non-standard entropy forms. 

The cosmological implications of the generalized mass-to-horizon entropy framework have been recently explored in Ref.~\cite{Gohar:2023lta}, where it was demonstrated that, for suitable choices of the model parameters, the resulting model shows excellent agreement with observational data, comparable to that of the standard $\Lambda$CDM model.
Moreover, by applying the gravity–thermodynamics conjecture, modified Friedmann equations were derived in Ref.~\cite{Basilakos:2025wwu}, leading to the emergence of an effective dark energy sector sourced by the additional terms arising from the generalized entropy expression. The associated dark energy equation-of-state parameter exhibits a dynamical behaviour, mimicking either quintessence or phantom energy at different redshifts, depending on the specific values of the entropic parameters. In addition, the ensuing cosmological scenario has been shown to be fully compatible with observational constraints from Supernova Type Ia, Cosmic Chronometers and Baryon Acoustic Oscillations datasets. 
These predictions can be directly compared with those of other recent dynamical and alternative dark energy models (see~\cite{Armendariz-Picon:2000nqq, Copeland:2006wr, Luciano:2023roh, Capolupo:2023fao, Carloni:2024ybx, Alfano:2024fzv}).

Building on the above premises, this work further investigates the impact of the generalized MHR and its associated cosmology on early Universe dynamics, with the aim of identifying potential observational signatures that could distinguish this model from the $\Lambda$CDM paradigm and other modified gravity scenarios. 
Specifically, we examine the formation and evolution of matter density perturbations, which serve as the primordial seeds of the large-scale structure observed in the present Universe, as well as the spectrum of primordial gravitational waves, generated by quantum fluctuations during the inflationary epoch. These signals offer a unique observational window into the pre-BBN evolution of the Universe, potentially constraining deviations from general relativity and standard thermodynamic assumptions. 

The structure of this work is as follows. In the next section, we begin by reviewing the conventional gravity-thermodynamics framework and then apply it to the context of generalized mass-to-horizon entropy, leading to the derivation of the modified Friedmann equations.  Section \ref{PGW} is devoted to the analysis of gravitational wave propagation in the early Universe and to constraining the model’s free parameter through its imprints on the associated spectrum. In Sec. \ref{SF}, we examine the implications for the growth of perturbations and structure formation. Conclusions and outlook are presented in Sec. \ref{Conc}. Throughout the manuscript, we adopt natural units.

\section{Modified cosmology through generalized mass-to-horizon
entropy}
\label{ModCos}
We begin our analysis by reviewing the application of the first law of thermodynamics within the framework of general relativity. This approach is then extended by introducing a modified MHR in place of the standard formulation. To fully explore the implications of this model, we generalize the study of \cite{Basilakos:2025wwu} to the case in which the perfect fluid permeating the Universe consists of both dust matter (i.e., cold dark matter and baryons) and radiation.

We carry out our discussion within the setting of a spatially flat Friedmann–Lema\^{\i}tre–Robertson–Walker (FLRW) background, described by the metric
\be
ds^2 = g_{\mu\nu}dx^{\mu}dx^{\nu} = \ell_{\alpha\beta}  dx^\alpha dx^\beta  +  \tilde r^2\left(d\theta^2  +  \sin^2\theta\, d\phi^2\right)\, ,
\label{FRW}
\ee
where $\tilde r=a(t)\hspace{0.2mm}r$, $x^0=t$, $x^1=r$, $\ell_{\alpha\beta}=\mathrm{diag}\left(-1,a^2\right)$ and $a(t)$ is the time-dependent scale factor.  

In this framework, the dynamical apparent horizon plays a central role in defining thermodynamic quantities. For the FLRW Universe, its radius is given by $\tilde{r}_A = 1/H$~\cite{Frolov:2002va,Cai:2005ra,Cai:2009qf}, where $H = \dot{a}/a$ denotes the Hubble parameter (with the dot indicating a derivative with respect to cosmic time). The associated temperature at the apparent horizon is typically taken to be the Hawking-like temperature \cite{Hawking:1975vcx}
\begin{equation}
\label{HaT}
   T_h = \frac{1}{2\pi \tilde{r}_A}\,,
\end{equation}
reflecting an analogy with black hole thermodynamics~\cite{Cai:2009qf,Padmanabhan:2009vy}. For our purposes, we adopt the assumption of a quasi-static cosmological expansion~\cite{Luciano:2023zrx}, which ensures that the temperature of the horizon remains well-defined throughout the evolution of the Universe. Additionally, we consider the cosmic fluid to be in thermal equilibrium with the apparent horizon, as a result of sustained interactions over cosmological timescales~\cite{Padmanabhan:2009vy,Frolov:2002va,Cai:2005ra,Izquierdo:2005ku,Akbar:2006kj}. This condition justifies the use of standard thermodynamic relations and allows us to avoid the complexities that arise in non-equilibrium treatments.

The next step involves assigning an entropy to the apparent horizon. Within the framework of general relativity, this is typically done using the standard Bekenstein–Hawking entropy derived from black hole thermodynamics, namely $S_{BH}=A/4$, where $A=4\pi \tilde r_A$ is the horizon area \cite{Bekenstein:1973ur,Bekenstein:1974ax}
.

Under the assumption that the Universe is filled with a perfect fluid, the energy-momentum tensor takes the standard form
\begin{equation}
\label{cont}
T_{\mu\nu} = (\rho + p)\, u_{\mu} u_{\nu} + p\, g_{\mu\nu} \,,
\end{equation}
where \( \rho \) is the energy density, \( p \) denotes the isotropic pressure and \( u^\mu \) is the four-velocity of the fluid. In this context, conservation of energy and momentum, $\nabla_\mu T^{\mu\nu} = 0$, yields the continuity equation
\begin{equation}
\dot{\rho} + 3H(\rho + p) = 0\,,
\end{equation}
which governs the evolution of \( \rho \) as the Universe undergoes expansion. The associated work density, arising from variations in the apparent horizon radius, is defined as \( \mathcal{W} = -\frac{1}{2} \, \text{Tr}(T^{\mu\nu}) = \frac{1}{2}(\rho - p) \), where the trace is computed with respect to the induced metric on the \((t, r)\) submanifold, that is, \( \text{Tr}(T^{\mu\nu}) = T^{\alpha\beta} h_{\alpha\beta} \).

At this stage, it is worth recalling that the gravity-thermodynamic conjecture suggests that Einstein’s equations can be derived from local thermodynamic relations applied to causal horizons. In a cosmological context, this leads to the striking result that the Friedmann equations can be derived from the first law of thermodynamics at the apparent horizon. To show this, let us consider the first law
\begin{equation}
dU = T_h\, dS - \mathcal{W}\, dV\,,
\label{14c}
\end{equation}
where \( dU \) denotes the increase in the internal energy of the Universe over an infinitesimal time interval \( dt \),  due to the change in the volume \( dV = 4\pi \tilde{r}_A^2 \, d\tilde{r}_A \) enclosed by the apparent horizon.  Observing that \( dU \) corresponds to a reduction in the total energy \( E = \rho V \) contained within that volume, i.e., \( dU = -dE \), Eq.~\eqref{14c} can be rearranged to give the second Friedmann equation
\begin{equation}
    \dot{H} = -4\pi \left(\rho + p\right)\,, 
    \label{F1}
\end{equation}
where we have used the approximation that the apparent horizon expands adiabatically \cite{Cai:2005ra,Sheykhi:2018dpn,Luciano:2023zrx,Luciano:2025hjn}.  

Inserting the matter conservation equation \eqref{cont} into Eq. \eqref{F1} and integrating, we obtain the first Friedmann equation
\begin{equation}
    H^2 = \frac{8\pi \rho}{3} +\frac{\Lambda}{3}\,,
    \label{F2}
\end{equation}
where the integration constant $\Lambda$ plays the role of the cosmological constant.

Therefore, applying the gravity-thermodynamic conjecture to the cosmic horizon yields the standard Friedmann equations. Clearly, as outlined above, extending this procedure to incorporate generalized entropy frameworks leads to modified forms of these equations. Such modifications naturally result in alternative cosmological models, in which the generalized entropic contributions effectively behave as a dark energy component \cite{Lymperis:2018iuz,Saridakis:2020lrg,Nojiri:2019skr,Hernandez-Almada:2021rjs,Dheepika:2022sio,Jizba:2022icu,Lambiase:2023ryq,Jizba:2024klq,Ebrahimi:2024zrk,Nojiri:2025gkq}. In what follows, we implement the gravity-thermodynamic approach by employing the generalized mass-to-horizon entropy introduced in \cite{Gohar:2023hnb,Gohar:2023lta}.

\subsection{Modified Friedmann equations through generalized MHR}

As discussed in Sec. \ref{Intro}, a generalized MHR has been proposed in \cite{Gohar:2023hnb,Gohar:2023lta} for application within the framework of entropic cosmology and scenarios based on the holographic principle. The key motivation behind this proposal stems from the observation that as long as the Clausius relation is employed to ensure thermodynamic consistency and a linear MHR of the form is assumed, then regardless of the entropy definition adopted, the resulting entropic force on cosmological horizons coincides with the standard one derived from Bekenstein entropy and Hawking temperature. As a result, such models inevitably face the same shortcomings as conventional Bekenstein–Hawking entropic frameworks in capturing the observed cosmic dynamics~\cite{Basilakos:2012ra,Basilakos:2014tha}.

These considerations have motivated the introduction of the generalized MHR \cite{Gohar:2023hnb,Gohar:2023lta}
\begin{equation}
    \label{GMHR}
    M=\gamma \frac{c^2}{G} L^n\,,
\end{equation}
where \( M \) and \( L \) represent the mass and the cosmological horizon of the system, respectively, \( \gamma \) is a positive parameter with dimensions $[L]^{1-n}$
and \( n \) is a non-negative constant
(we have temporarily restored the fundamental constants $c$ and $G$ for consistency with Refs.~\cite{Gohar:2023hnb,Gohar:2023lta}). This relation is the minimal generalization that allows one to preserve Hawking’s temperature and the Clausius relation consistently on the cosmological horizon.
Interestingly, for suitable choices of the entropic parameters, the resulting cosmological model shows excellent agreement with observational data, demonstrating its viability as an alternative to the standard cosmological paradigm and thus providing new fundamental support for the physical origin and nature of the cosmological constant.

By combining Eq.~\eqref{GMHR} with the Clausius relation and employing the Hawking temperature defined in Eq.~\eqref{HaT}, it is now possible to derive a generalized expression for the entropy of the form \cite{Gohar:2023hnb,Gohar:2023lta}
\begin{equation}
\label{GMHE}
    S_n=\gamma\hspace{0.2mm}\frac{2n}{n+1}\hspace{0.2mm}\tilde r_A^{n-1}\hspace{0.2mm}S_{BH}\,,
\end{equation}
where $S_{BH}$ denotes the Bekenstein–Hawking entropy and the apparent horizon $\tilde{r}_A$ has been taken as the characteristic length scale $L$.

Based on Eqs.~\eqref{GMHR} and~\eqref{GMHE}, it is worthwhile to elaborate on the physical meaning of the parameters $n$ and $\gamma$, and to outline limiting cases that establish connections with well-defined gravitational and cosmological frameworks. From Eq.~\eqref{GMHR}, it is clear that the exponent $n$ controls the scaling of the horizon mass with the horizon radius. In turn, from Eq.~\eqref{GMHE} we infer that $n>1$ corresponds to a super-extensive regime, in which the modified entropy grows faster than the Bekenstein--Hawking area law, while $n<1$ would instead signal a sub-extensive scaling.
Furthermore, as discussed in~\cite{Gohar:2023lta} (see also Eq.~\eqref{rhode} below), this parameter determines how the entropic energy density associated with the generalized MHR
evolves with the Hubble rate, namely $\rho_e \propto H^{3-n}$. On the other hand, the parameter $\gamma$ acts as a coupling-like constant. Physically, it quantifies how efficiently the degrees of freedom associated with the horizon are converted into an effective bulk energy density~\cite{Gohar:2023hnb,Gohar:2023lta}.

An especially relevant case of the extended model \eqref{GMHR} arises for $n=3$, where the entropy scales as $S_n \propto L^4$, while the mass grows proportionally to the volume ($M \propto L^3$). In this scenario, the entropic density $\rho_e$ remains constant, thereby reproducing a cosmological constant-like behavior. Another noteworthy case is $n=2$, which corresponds to a mass scaling with the horizon surface ($M \propto L^{2}$), while the associated entropy becomes extensive in three dimensions and scales as $S_{n} \propto L^{3}$~\cite{Gohar:2023hnb,Gohar:2023lta}. Finally, the limiting case $n=1$ with $\gamma=1/2$ reproduces the linear Misner--Sharp mass in a spherically symmetric spacetime~\cite{Gong:2007md}, defined with respect to the apparent horizon. By contrast, setting both $n$ and $\gamma$ to unity yields the standard entropy-area law, with the entropy density scaling as $H^{2}$, thereby recovering the linear MHR usually assumed in analogy with black holes together with the Bekenstein entropy $S_{BH}$.

In general, since any plausible deviations from the standard entropic framework are expected to be relatively small, in the following analysis we restrict our attention to perturbative departures from $n=1$. This assumption is consistent with the observational constraints reported in~\cite{Gohar:2023lta,Basilakos:2025wwu}. Moreover, for the study presented in Sec.~\ref{PGW} we adopt $\gamma=1$, in line with the setting of~\cite{Basilakos:2025wwu}. From a theoretical standpoint, because $\gamma$ enters as a purely multiplicative factor in Eq.~\eqref{GMHE}, deviations from unity are expected to be either negligible or at least subdominant compared to the effects induced by non-standard values of the exponent $n$. On the observational side, this choice is further supported by the recent results of~\cite{Luciano:2025ovj}, which constrain $\gamma$ to values close to unity.

The generalized mass-to-horizon entropy relation \eqref{GMHE} has been adopted as the starting point for deriving a modified cosmological scenario in Ref.~\cite{Basilakos:2025wwu}. In particular, by using the gravity–thermodynamics conjecture and following the same steps outlined above, one arrives at the modified Friedmann equations
\begin{eqnarray}
\label{FM1}
    H^2&=&\frac{8\pi}{3}\left(\rho+\rho_{DE}\right),\\[2mm]
    \dot H&=&-4\pi \left(\rho+p+\rho_{DE}+p_{DE}\right),
    \label{FM2}
\end{eqnarray}
where the effect of the generalized entropy manifests itself through the emergence of an effective dark energy component, characterized by an energy density and pressure given by
\begin{eqnarray}
\label{rhode}
    \rho_{DE}&=&\frac{3}{8\pi}\left[\frac{\Lambda}{3}+H^2-\frac{2\gamma n}{3-n}\hspace{0.2mm}H^{3-n}\right]\,\\[2mm]
    p_{DE}&=&-\frac{1}{8\pi}\left[\Lambda + \left(2\dot H+3H^2\right)
    -2\gamma n H^{1-n}\left(\dot H+\frac{3}{3-n}H^2\right)
    \right],
\end{eqnarray}
respectively. Once again, it can be verified that by setting $n = \gamma = 1$, the standard scenario is recovered, as this leads to $\rho_{DE} = -p_{DE}=\Lambda/(8\pi)$.
Therefore, we emphasize that the appearance of a dark energy term is not due to the explicit inclusion of such a component in the Universe's energy budget, but rather arises naturally from the generalizations introduced in Eqs.~\eqref{GMHR} and \eqref{GMHE}.

The Friedmann equations derived above can be further manipulated by introducing the fractional energy densities $\Omega_i \equiv 8\pi\rho_i/(3H^2)$ (the index $i = m, r$ corresponds  to matter and radiation, respectively) and $\Omega_{{DE}} \equiv {8\pi\rho_{{DE}}}/(3H^2)$. From Eq. \eqref{FM1}, we obtain
\begin{equation}
\label{cond1}
\Omega_{m}+\Omega_r+\Omega_{DE}=1\,.    
\end{equation}
Denoting the present-day energy densities of pressureless matter and radiation by \( \rho_{m0} \) and \( \rho_{r0} \), and assuming their standard evolution laws \( \rho_m(a) = \rho_{m0} / a^3 \) and \( \rho_r(a) = \rho_{r0} / a^4 \), we are led to\footnote{Throughout the manuscript, we adopt the standard convention in which the subscript “0” denotes the present-day value of a given quantity.}
\begin{equation}
\label{Hquad2}
    H^2(a)=\frac{H_0^2}{1-\Omega_{DE}(a)}\left(\frac{\Omega_{m0}}{a^3}+\frac{\Omega_{r0}}{a^4}
    \right).
\end{equation}
It is convenient to use the redshift $z$ as the independent variable, defined by the relation $1 + z = 1 / a$. By replacing Eq. \eqref{rhode} into Eq. \eqref{Hquad2}, we acquire 
\begin{equation}
\label{Omde}
\Omega_{DE}(z)=1\,-\,\Bigg\{\frac{3-n}{2\gamma n}\left(H_0\sqrt{\Omega_{m0}+\Omega_{r0}\left(1+z\right)}\right)^{n-1}\left(1+z\right)^{\frac{3(n-1)}{2}}\bigg\{1\,+\,\frac{\Lambda}{3H_0^2\left[\Omega_{m0}+\Omega_{r0}\left(1+z\right)\right]\left(1+z\right)^3}\bigg\}
    \Bigg\}^{\frac{2}{n-3}}\,.
\end{equation}
Applying this relation at the present time allows us to express the cosmological constant as
\begin{equation}
    \Lambda=\frac{6\gamma n}{3-n}\hspace{0.2mm} H_0^{3-n}-3H_0^2\left(\Omega_{m0}+\Omega_{r0}\right),
    \label{Lambda}
\end{equation}
where we have used the condition \eqref{cond1}.

The late-time cosmological implications of the model described by Eq.~\eqref{Omde} have been recently examined in Ref.~\cite{Basilakos:2025wwu}, where it was demonstrated that the Universe undergoes the standard thermal evolution, featuring the successive dominance of matter and dark energy epochs. Moreover, depending on the specific values of the entropic parameters, the dark energy equation-of-state parameter may either reside in the phantom regime at high redshifts and transition into the quintessence regime at low redshifts, or alternatively, remain within the quintessence regime at early times and subsequently cross the phantom divide at later stages. An observational analysis was also conducted using Supernova Type Ia, Cosmic Chronometers and Baryon Acoustic Oscillations datasets, revealing that the model exhibits good consistency with current observational data.

Here, we aim to further investigate the predictions of this extended cosmological model. For the purposes of the following analysis, it is convenient to work with the modified Hubble rate expression. To this end, substituting Eq.~\eqref{Omde} back into Eq.~\eqref{Hquad2}, we infer
\begin{eqnarray}
\nonumber
H^2(z)&=&2^{\frac{2}{n-3}}H_0^2\left[\Omega_{m0}\left(1+z\right)^3+\Omega_{r0}\left(1+z\right)^4\right]\\[2mm]
    &&\times\,\Bigg\{\frac{3-n}{\gamma n}\left(H_0\sqrt{\Omega_{m0}+\Omega_{r0}\left(1+z\right)}\right)^{n-1}\left(1+z\right)^{\frac{3(n-1)}{2}}\bigg\{1+\frac{\Lambda}{3H_0^2\left[\Omega_{m0}+\Omega_{r0}\left(1+z\right)\right]\left(1+z\right)^3}\bigg\}
    \Bigg\}^{\frac{2}{3-n}}\,.
    \label{ModHubrate}
\end{eqnarray}
This expression reproduces the standard $\Lambda$CDM evolution in the limit $n = \gamma = 1$, as expected. In addition, due to condition~\eqref{Lambda}, it follows that $H(z = 0) = H_0$, independently of the specific values of the entropic parameters. 

\begin{figure}[t]
\begin{center}
\includegraphics[width=9cm]{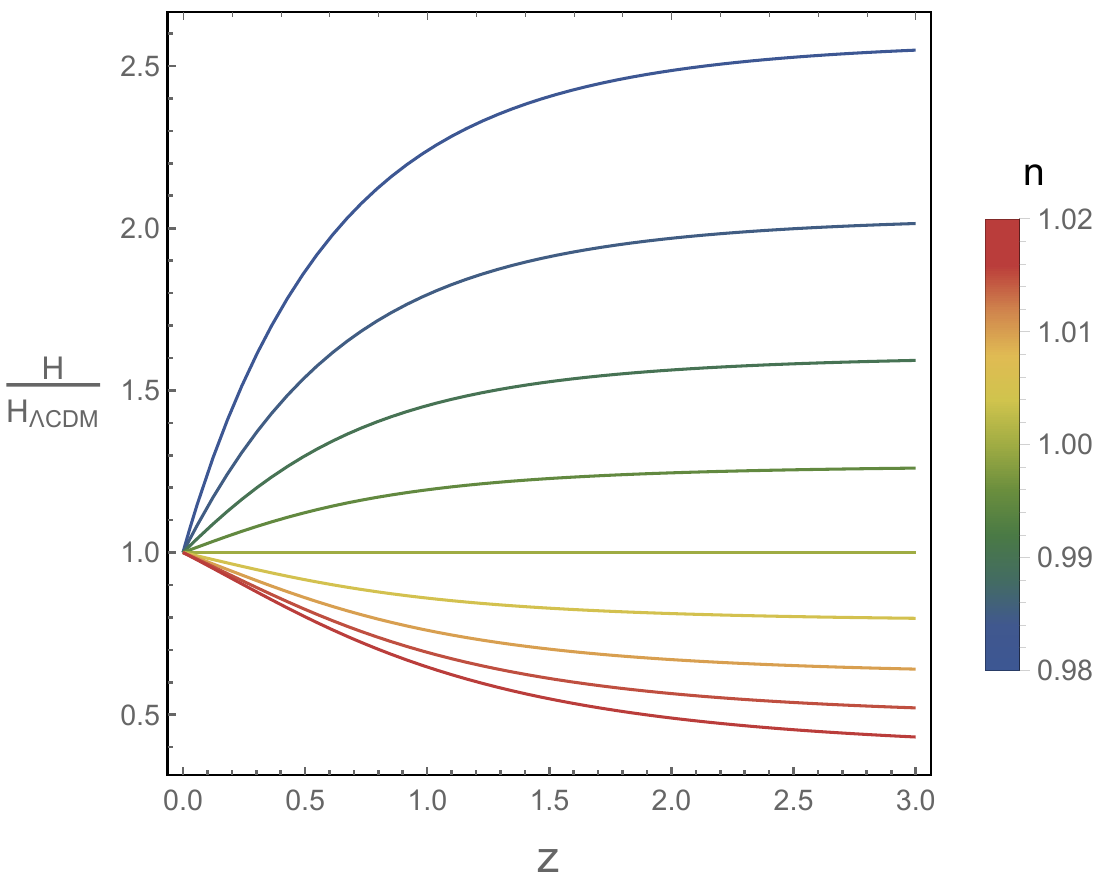}
\caption{Plot of the ratio \( H(z)/H_{\Lambda \mathrm{CDM}} \) versus $z$, for different values of \( n \). We used the modified expression of the Hubble rate in Eq.~\eqref{ModHubrate} and set $\Omega_{m,0}\simeq 0.3$,  $\Omega_{r,0}\simeq 10^{-5}$.}
\label{Fig0}
\end{center}
\end{figure}

In Fig.~\ref{Fig0}, we plot the ratio \( H(z)/H_{\Lambda \mathrm{CDM}} \) as a function of redshift \( z \). Since we are primarily interested in the effects of the entropic exponent \( n \), which we expect to introduce the main differences in the cosmic evolution, we fix \( \gamma = 1 \) and consider various values of \( n \), as in Ref.~\cite{Basilakos:2025wwu}. It is evident that for \( n < 1 \), corresponding to a generalized mass-to-horizon entropy that grows more slowly than the Bekenstein-Hawking entropy (see Eq.~\eqref{GMHE}), the modified Hubble rate exceeds its standard counterpart (\( n = 1 \)) at early times. In contrast, for \( n > 1 \), the opposite behavior is observed, with the Hubble rate falling below the standard prediction. This behavior highlights the non-trivial role of the entropic index \( n \) in cosmic evolution, as it significantly affects the expansion dynamics.

From a thermodynamic perspective, this behavior can be intuitively understood as follows: when the entropy increases more rapidly than in the Bekenstein–Hawking case (\( n > 1 \)), a larger fraction of the total energy content of the Universe becomes effectively stored in the gravitational degrees of freedom associated with the horizon. Consequently, less energy remains available to drive the cosmic expansion, resulting in a slower growth of the scale factor and, therefore, a suppressed Hubble rate compared to the standard scenario. Naturally, the opposite occurs in the case \( n < 1 \).

In the following sections, we explore how the result \eqref{ModHubrate} impacts the spectrum of primordial gravitational waves, as well as its implications for the growth of matter perturbations and the formation of cosmic structures.

\section{Primordial Gravitational Wave Signals}
\label{PGW}

Primordial Gravitational Waves (PGWs) are thought to preserve signatures of quantum fluctuations and potential phase transitions that took place during the inflationary era of the early Universe~\cite{Maggiore:1999vm,Saikawa:2018rcs,Giare:2022wxq}. Detecting such imprints would be of significant importance, as it would offer a unique opportunity to explore the Universe's history prior to Big Bang Nucleosynthesis (BBN), including phases such as reheating, the hadronic and quark epochs and possible early non-standard phases dominated by either matter or kination. Moreover, since general relativity is expected to be modified by quantum corrections in the ultraviolet regime, the pre-BBN era serves as an ideal setting for testing alternative theories of gravity applicable to the early Universe.

In scenarios beyond the conventional single-field slow-roll inflation, the production of GWs can result in signals that may be detectable at scales smaller than those probed by the Cosmic Microwave Background (CMB). In this section, we compute the spectrum of primordial gravitational waves within the framework of the generalized cosmological model developed in Sec.~\ref{ModCos} and compare it with the predictions of the standard cosmological scenario. The resulting observational signatures and their potential detectability will be examined in detail. For this purpose, we follow the formalism developed in Ref.~\cite{Watanabe:2006qe} and recently employed in Refs.~\cite{Bernal:2020ywq, Barman:2023ktz,Maity:2024cpq,Jusufi:2024utf,Luciano:2024mcn}.

\subsection{PGW in standard cosmology}
\label{PGWst}

In the linearized regime, GWs are treated as perturbations of the metric on a curved spacetime background. We focus on tensor perturbations propagating over a homogeneous, isotropic and spatially flat background spacetime. In this setting, it is appropriate to impose $h_{00} = h_{0i} = 0$, effectively eliminating time and mixed components of the perturbations. Furthermore, we adopt the transverse-traceless (TT) gauge, characterized by the conditions $\partial_i h_{ij} = 0$ and $h^i_i = 0$ (latin
indices run over the three spatial coordinates). Under these assumptions, the evolution of tensor perturbations at first order is governed by the following equation \cite{Watanabe:2006qe}
\begin{equation}
\label{hdyn}
\ddot h_{ij}+ 3H\dot h_{ij}  -  \frac{\nabla^2}{a^2}h_{ij}= 16\pi G\hspace{0.3mm} \Pi_{ij}^{TT}\,,
\end{equation}
where $\Pi_{ij}^{TT}$ is the $TT$ anisotropic part of the stress tensor
\be
\Pi_{ij} \ = \ \frac{T_{ij}-p\hspace{0.3mm} g_{ij}}{a^2}\,,
\label{TPE}
\ee
and $T_{ij}, g_{ij}$ and $p$ are the stress-energy tensor, the metric tensor and the background pressure, respectively. 

We remark that Eq.~\eqref{hdyn} is the standard wave equation for tensor perturbations propagating on a FRW background. Its derivation follows within the conventional Lagrangian (field-theoretic) formalism of general relativity. 
While this equation is independent of the thermodynamic arguments used to motivate the generalized MHR, it nevertheless serves as the standard starting point for describing GWs in extended cosmological backgrounds too~\cite{Bernal:2020ywq}.
In the following analysis, our strategy is to adopt Eq.~\eqref{hdyn} in the standard form, and then investigate how the background dynamics implied by the entropic cosmological model affects the evolution and observational features of the primordial tensor spectrum. In other words, the thermodynamic framework provides the modified background expansion, while the perturbations are consistently treated within the conventional field-theoretic formalism. This separation ensures that the results can be directly compared with those obtained in the standard cosmological scenario and with the sensitivities of forthcoming GW observatories. A complete treatment would eventually require extending the framework to include perturbations consistently within the entropic approach. This important development, however, goes beyond the scope of the present work and will be addressed in a future investigation.

Furthermore, it is important to note that, although the background spacetime is homogeneous and isotropic, the presence of $\Pi_{ij}^{TT}$ in Eq. \eqref{hdyn}  accounts for the possible existence of anisotropic stress perturbations in the cosmic fluid or fields. This term encapsulates the transverse-traceless part of the stress-energy tensor perturbations, which can arise from various physical sources such as free-streaming relativistic particles (e.g., neutrinos), magnetic fields or other non-perfect fluid components. In the absence of such sources, the evolution of GWs would be governed solely by the homogeneous part of Eq.~\eqref{hdyn}, describing their free propagation with damping due to cosmic expansion. However, when anisotropic stresses are present, they act as a source term capable of generating or modifying gravitational wave amplitudes. Therefore, $\Pi_{ij}^{TT}$ plays a crucial role in the dynamics of tensor perturbations, encoding the imprint of microphysical processes and matter content on the evolution of GWs in the early Universe.

GW signals are typically classified according to their generation mechanisms into three main categories: inflationary, cosmological and astrophysical sources, with their characteristic frequencies strongly depending on the underlying production process. In the present work, we focus on GWs within the frequency interval $[10^{-11},10^3]\,\mathrm{Hz}$. This frequency range is particularly relevant, since it is expected to be fully
tested by current and upcoming GW observatories. 

In order to solve Eq.~\eqref{hdyn}, it is convenient to work in Fourier space, where the tensor perturbations can be expressed as~\cite{Watanabe:2006qe}
\begin{equation}
h_{ij}(t, \vec{x}) = \sum_{\lambda} \int \frac{d^3k}{(2\pi)^3} \, h^\lambda(t, \vec{k}) \, \epsilon^\lambda_{ij}(\vec{k}) \, e^{i \vec{k} \cdot \vec{x}}\,,
\end{equation}
where $\epsilon^\lambda_{ij}$ denotes the spin-2 polarization tensor, which satisfies the orthonormality condition $\sum_{ij} \epsilon^\lambda_{ij} \epsilon^{\lambda'*}_{ij} = 2 \delta^{\lambda \lambda'}$, with $\lambda = +, \times$ labeling the two independent polarization states of GWs.

The tensor perturbation $h^\lambda(t, \vec{k})$ can be decomposed as
\begin{equation}
h^\lambda(t, \vec{k}) = h_{\mathrm{prim}}^\lambda(\vec{k}) \, X(t, k)\,,
\end{equation}
where $k = |\vec{k}|$, $X(t, k)$ is the transfer function describing the time evolution of the perturbation and $h_{\mathrm{prim}}^\lambda(\vec{k})$ represents the primordial amplitude of the tensor perturbations. Within this parametrization, the tensor power spectrum reads~\cite{Bernal:2020ywq}
\be
\mathcal{P}_T(k) =  \frac{k^3}{\pi^2}\sum_\lambda\Big|h^\lambda_{\mathrm{prim}}(\vec k)\Big|^2  =  \frac{2}{\pi^2}\hspace{0.3mm}G\hspace{0.3mm} H^2\Big|_{k=aH}\,.
\ee
Since the spectrum depends explicitly on \( H^2 \), it is evident that any deviation from the background expansion predicted by the \( \Lambda \)CDM model is expected to leave a detectable imprint in the PGW spectrum. 

On the other hand, Eq.~\eqref{hdyn} reduces to a differential equation analogous to that of a damped harmonic oscillator, i.e., 
\begin{equation}
\label{Xd}
X'' + 2 \frac{a'}{a} X' + k^2 X = 0\,,
\end{equation}
where primes denote derivatives with respect to conformal time $\tau$, defined through $d\tau = dt / a$.

The relic density of PGWs arising from first-order tensor perturbations within the standard cosmological framework is defined by~\cite{Watanabe:2006qe}
\begin{equation}
\Omega_{\mathrm{GW}}(\tau, k) = \frac{\left[ X'(\tau, k) \right]^2}{12 a^2(\tau) H^2(\tau)} \, \mathcal{P}_T(k) \simeq \left[ \frac{a_{\mathrm{hc}}}{a(\tau)} \right]^4 \left[ \frac{H_{\mathrm{hc}}}{H(\tau)} \right]^2 \frac{\mathcal{P}_T(k)}{24}\,,
\label{Ttps}
\end{equation}
where, in the second step, we have averaged over oscillation periods, which implies 
\be
X'(\tau,k)\simeq  k\hspace{0.2mm} X(\tau,k) \simeq  \frac{k\hspace{0.3mm} a_{\mathrm{hc}}}{\sqrt{2}a(\tau)}\ \simeq \
\frac{a^2_{\mathrm{hc}}\hspace{0.3mm}H_{\mathrm{hc}}}{\sqrt{2}a(\tau)}\,,
\ee
with $k=2\pi f=a_{\mathrm{hc}}H_{\mathrm{hc}}$ at the horizon crossing. We note that this approximation is evaluated at the horizon crossing because it corresponds to the moment when a given mode with wavenumber \( k \) re-enters the Hubble radius, that is, when its physical wavelength matches the Hubble horizon size. At this stage, the dynamics of the perturbation become sensitive to the background expansion and its amplitude begins to decay due to the cosmic damping. Consequently, the horizon-crossing time sets a natural reference point for the subsequent evolution of the gravitational wave mode.

By introducing the reduced Hubble constant $h$, the relic density of PGWs can be expressed as
\begin{equation}
\Omega_{\mathrm{GW}}(\tau_0, k) \, h^2 \simeq \left[ \frac{g_*(T_{\mathrm{hc}})}{2} \right] \left[ \frac{g_{*s}(T_0)}{g_{*s}(T_{\mathrm{hc}})} \right]^{4/3} \frac{ \mathcal{P}_T(k) \, \Omega_r(T_0) \, h^2 }{24}\,,
\label{Spt}
\end{equation}
where $g_*(T)=\frac{\pi^2}{30}g_*(T)T^4$ and $g_{*s}(T)=\frac{2\pi^2}{45}g_{*s}(T)T^3$ are the effective numbers of relativistic degrees of freedom that contribute to the radiation energy
density $\rho$ and entropy density $s$, respectively. Furthermore, we have reparametrized the evolution in terms of the temperature, which can be related to the redshift through the standard relation $1+z = \dfrac{T}{T_0}$, where where $T_0\simeq 3\,\mathrm{K}$ is the average temperature of the observable Universe at present time.

The scale-dependence of the tensor power spectrum is given by \cite{Watanabe:2006qe,Bernal:2020ywq}
\begin{equation}
\mathcal{P}_T(k)  =  A_T \left( \frac{k}{\tilde{k}} \right)^{n_T} \,,
\end{equation}
where \( n_T \) denotes the tensor spectral index and \( \tilde{k} = 0.05\,\mathrm{Mpc}^{-1} \) represents a reference pivot scale commonly used in cosmological analyses. The amplitude \( A_T \) of the tensor modes is connected to the scalar perturbation amplitude \( A_S \) through the tensor-to-scalar ratio \( r \), such that \( A_T = r A_S \).

The relic density Eq.~\eqref{Spt} is displayed as a function of the frequency \( f \) in Fig.~\ref{Fig1} (blue solid line), under the assumption of a scale-invariant primordial tensor power spectrum (\( n_T \simeq 0 \)) and 
normalized according to the amplitude of scalar perturbations measured by Planck. Specifically, the scalar amplitude at the pivot scale  $\tilde k$ is constrained to be \( \ln (10^{10} A_S) = 3.044 \pm 0.014 \), corresponding to \( A_S \simeq 2.1 \times 10^{-9} \)~\cite{Planck:2018vyg}.
The shaded areas correspond to the expected sensitivity ranges of several present and upcoming GW observatories~\cite{Breitbach:2018ddu}, including the LISA mission~\cite{amaro2017laser}, the Einstein Telescope (ET)~\cite{Sathyaprakash:2012jk}, the proposed Big Bang Observer (BBO)~\cite{Crowder:2005nr} and the Square Kilometre Array (SKA)~\cite{Janssen:2014dka}. Moreover, constraints from Big Bang Nucleosynthesis (BBN) are shown, derived from observational limits on the effective number of neutrino species~\cite{Boyle:2007zx,Stewart:2007fu}. The gray regions indicate the parameter space excluded by current Pulsar timing array (PTA)~\cite{KAGRA:2021kbb} and LIGO~\cite{Shannon:2015ect} data, respectively.

\subsection{PGW in modified cosmology}
\label{PGWmod}
We now proceed to analyze the impact of the modified dynamics \eqref{ModHubrate} on the PGW spectrum. 
We assume that modifications to gravity predominantly affect the evolution of the Universe at the background level, meaning that all corrections within our extended cosmological framework are effectively captured by the modified Hubble parameter. This assumption remains valid as long as deviations from general relativity are small, which is precisely the regime considered in this work. In a broader context, however, modifications to gravity can also influence linear perturbations, potentially affecting the transfer functions and the resulting power spectra of scalar and tensor modes produced during inflation. 
Moreover, additional imprints may also emerge from modifications in the formation and dynamics of primordial black holes. Indeed, the generation of such objects is associated with enhanced curvature perturbations on small scales, which can act as sources of a stochastic GW background through second-order effects. Such mechanisms could leave distinctive signatures in the PGW spectrum, potentially providing complementary probes of modified gravity and early-Universe physics~\cite{Carr:2020gox, Sasaki:2018dmp,Basilakos:2023xof}. A thorough investigation of such effects, including their impact on the perturbation dynamics, lies beyond the scope of the present study and is left for future work.

To incorporate the effects of the generalized mass-to-horizon entropy, we observe that Eq.~\eqref{Ttps} can be rewritten in the equivalent form 
\begin{eqnarray}
\nonumber
 \Omega_{\mathrm{GW}}(\tau,k)&=& \!\left[\frac{a_{\mathrm{hc}}}{a(\tau)}\right]^4\left[\frac{H_{\mathrm{hc}}}{H_{\mathrm{GR}}(\tau)}\right]^2\left[\frac{H_{\mathrm{GR}}(\tau)}{H(\tau)}\right]^2\frac{\mathcal{P}_T(k)}{24} \\[2mm]
 &=&\Omega^{\mathrm{GR}}_{\mathrm{GW}}(\tau,k)\left[\frac{H_{\mathrm{GR}}(\tau)}{H(\tau)}\right]^2\left[ \frac{a_{\mathrm{hc}}}{a_{\mathrm{hc}}^{\mathrm{GR}}}\right]^4 \left[ \frac{a^{\mathrm{GR}}(\tau)}{a(\tau)}\right]^4\left[ \frac{H_{\mathrm{hc}}}{H_{\mathrm{hc}}^{\mathrm{GR}}}\right]^2   \,.
\label{eq:PGWBar0}
\end{eqnarray}
Here, the subscript/superscript ``GR'' refers to quantities evaluated within the framework of standard general relativity. For example, \( \Omega^{\mathrm{GR}}_{\mathrm{GW}}(\tau, k) \) denotes the predicted PGW energy density according to Einstein's theory, which corresponds to the expression given in Eq.~(\ref{Ttps}).

\begin{figure}[t]
\begin{center}
\includegraphics[width=18 cm]{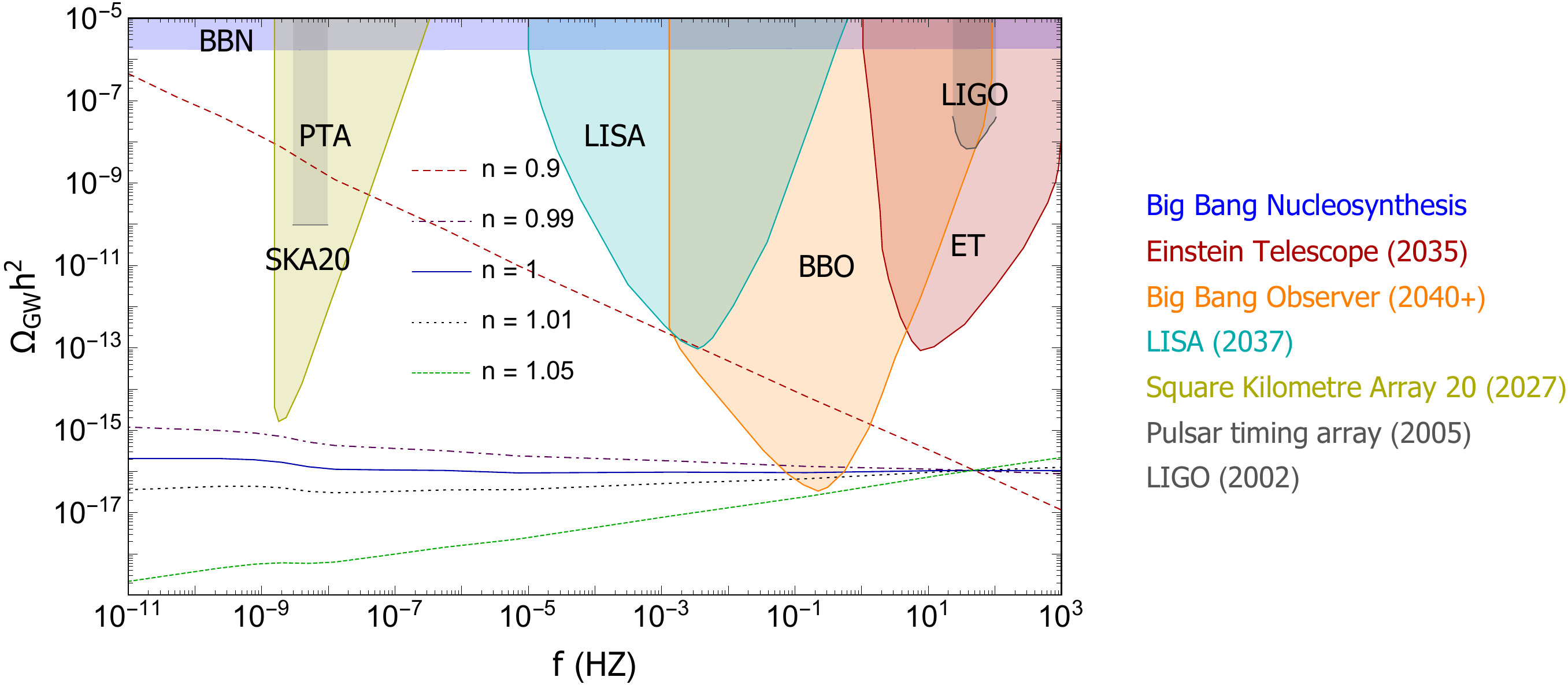}
\caption{Plot of the PGW spectrum versus the frequency $f$ for $n_T=0$, $A_S\simeq2.1\times10^{-9}$ and  different value of the entropic parameter $n$, while keeping $\gamma$ fixed to its value in the standard theory. The shaded regions represent the forecasted sensitivity ranges of various GW detectors~\cite{Breitbach:2018ddu}, which are listed on the right together with their expected launch years.  In gray we show the regions excluded by  PTA~\cite{KAGRA:2021kbb} and LIGO~\cite{Shannon:2015ect}.}
\label{Fig1}
\end{center}
\end{figure}

By exploiting the condition that, at the present time, the modified cosmological scenario under consideration coincides with standard general relativity, we can finally rewrite Eq. \eqref{eq:PGWBar0} as
\begin{equation}
 \Omega_{\mathrm{GW}}(\tau_0,k)
= \Omega^{\mathrm{GR}}_{\mathrm{GW}}(\tau_0,k)\left[ \frac{a_{\mathrm{hc}}}{a_{\mathrm{hc}}^{\mathrm{GR}}}\right]^4\left[ \frac{H_{\mathrm{hc}}}{H_{\mathrm{hc}}^{\mathrm{GR}}}\right]^2   \,. \label{eq:PGWBar} 
\end{equation}
Figure~\ref{Fig1} shows the PGW spectrum as a function of the frequency \( f \), computed for different values of the entropic parameter \( n \) using Eq.~\eqref{eq:PGWBar}. Two distinct cases must be primarily distinguished in the analysis:
\begin{itemize}
\item[-] for \( n > 1 \), the PGW spectrum becomes increasingly suppressed at low frequencies as \( n \) increases. This behavior is primarily due to the fact that, as shown in Fig.~\ref{Fig0}, the modified Hubble rate is lower than its counterpart in general relativity, resulting in a reduced relic density \eqref{eq:PGWBar} relative to the conventional scenario.
Should PGWs be detected, particularly given the experimental sensitivity of BBO, it would be possible to constrain deviations from the linear MHR up to \( n - 1 < 0.05 \) (green dotted line)\footnote{It is worth noting that this bound only provides a forecast based on the expected sensitivity of BBO, which is a proposed successor to the LISA mission that is not yet operational. Clearly, a quantitatively more explicit constraint, with well-defined statistical uncertainties, will only be achievable once observational data from future GW observatories become available.}. Indeed, larger values of \( n \) would produce an even more suppressed spectrum, rendering such signals undetectable by upcoming GW observatories.

On the other hand, the absence of detectable PGW signatures would further suggest that the classical cosmological model may need to be extended to maintain phenomenological consistency. From the perspective of the generalized MHR \eqref{GMHR}, such an extension could be encoded in the form of a generalized entropy-area relation, involving deviations of the order $n-1\gtrsim0.05$. In passing, we note that this latter scenario appears to be supported by the findings of Ref.~\cite{Basilakos:2025wwu}, where the analysis yields \( n \simeq 1.1 \) at the \( 1\sigma \) confidence level,  relying on late-time measurements from CC, SnIa and BAO data.

\item[-] Conversely, for \( n < 1 \), the PGW spectrum would be enhanced relative to the prediction of general relativity, consistently with the faster growth of the Hubble rate observed in this regime (see Fig.~\ref{Fig0}). If this were indeed the case in Nature, signatures of PGWs could potentially be detected by LISA or SKA20 (in addition to BBO), even at frequencies below \( 10^3 \,\mathrm{Hz} \). {Interestingly, by taking into account the region excluded by the PTA datasets in the nanohertz band~\cite{Reardon:2023gzh} (see Fig. \ref{Fig1}) and requiring that the modified spectrum does not intersect this region, one can infer a lower bound on the entropic index, namely 
$n > n_-= 0.884^{+0.002}_{-0.001}$ (1$\sigma$ CL)}.
\end{itemize}

It is worth comparing the present result with the modification to the PGW spectrum induced by the \(\delta\)-Tsallis entropy \cite{Jizba:2024klq}, which also predicts a power-law deformation of the horizon entropy, although motivated by different considerations related to non-extensive statistical mechanics \cite{Tsallis:2013}. In that context, it is shown that for sub-extensive scaling of the holographic degrees of freedom, corresponding to \( \delta < 1 \), the PGW spectrum is enhanced compared to general relativity. Conversely, the opposite behavior arises in the case of super-extensive scaling, i.e., \( \delta > 1 \), where the spectrum is suppressed relative to the standard prediction.
These features are consistent with the findings of the present work, taking into account that, as discussed in \cite{Gohar:2023lta}, the correspondence between the \( \delta \)-Tsallis entropy and the generalized mass-to-horizon entropy relation is established for \( n + 1 = 2 \delta \).

\subsection{Comparison with other generalized gravity frameworks}

{We note that deviations in the PGW spectrum may also originate from alternative sources. It is therefore crucial to carefully distinguish such effects from the specific signatures predicted by the cosmological model investigated in this work. Indeed, such anomalies are typically difficult to explain within the conventional FRW cosmological framework and often necessitate appropriate modifications to either the matter content or the gravitational sector of Einstein’s equations.}

{In this respect, it is important to acknowledge that our generalized MHR framework is, to some extent, 
formally degenerate with other non-extensive entropy proposals, such as the Tsallis--Cirto and Barrow entropies, 
since they share the same functional form of power-law deformations of the Bekenstein--Hawking entropy. This degeneracy 
makes it challenging, at the observational level, to disentangle the signatures of the generalized MHR scenario from 
those of Tsallis-Barrow--like entropies \cite{Jizba:2024klq}. }

{However, when compared with other classes of models, distinctive differences emerge. 
For example, the presence of a massive axion field can lead to the emergence of a distinctive triangular peak in the PGW spectrum \cite{Ax1}, with the characteristics of this feature encoding valuable information about the underlying axion potential. Similarly, it has been shown in \cite{Ax2} that such a triangular signature may also arise from axion kination scenarios originating during the inflationary epoch. These
imprints differ qualitatively from the smooth power-law deformations predicted
by entropy-based models.}

{On the other hand, the impact of modified gravity theories has been extensively investigated across various frameworks; for a comprehensive review, see \cite{Odintsov:2022cbm}. In particular, scalar-tensor theories and extra-dimensional models have been analyzed in \cite{Bernal:2020ywq} through the parametrization \( H(T) = A(T) H_{\mathrm{GR}}(T) \) for the Hubble expansion rate, where \( A(T) \) represents an amplification factor commonly expressed as \( A(T) = 1 + \eta \left( T / T_* \right)^\nu \). Here, \( T_* \) denotes the characteristic temperature at which deviations from general relativity become significant, while \( \eta \) and \( \nu \) are dimensionless parameters determined by the specific cosmological model. Notably, the parameter \( \nu \) plays a crucial role, as varying its value leads to different theoretical predictions for the shape of the PGW spectrum across diverse cosmological scenarios. Specifically, for \( \nu > 0 \), the modified PGW spectrum exhibits a blue-tilted behavior relative to the standard prediction, leading to significant deviations from general relativity at high frequencies. This behavior is characteristic of scenarios such as Randall–Sundrum type II brane cosmology (\( \nu = 2 \)) and kination models (\( \nu = 1 \)). In contrast, the case \( \nu = 0 \) corresponds to a constant amplification of the Hubble rate, with \( A = 1 + \eta \), resulting in an overall frequency-independent enhancement of the PGW spectrum. A similar effect also arises in the presence of a large number of additional relativistic degrees of freedom within the thermal plasma \cite{Catena:2009tm}. Finally, for \( \nu < 0 \), the modification leads to a localized enhancement of the spectrum, producing a characteristic bump near the reheating frequency.}

In comparison, the modified PGW spectrum derived in our analysis exhibits distinctive anomalies arising from the generalized MHR, which become particularly significant at low frequencies. Specifically, for $n>1$ the spectrum displays a smooth decrease with respect to the GR prediction, whereas for $n<1$ an enhancement emerges, which in principle could allow the detection of PGW signatures also by LISA and SKA20 (in addition to BBO), even at frequencies below $10^{3}\,\mathrm{Hz}$.
In light of the above discussion, it can be inferred that the profile depicted in Fig.~\ref{Fig1} may serve as a distinctive signature of the generalized mass-to-horizon entropy relation, and, more broadly, of power-law deformations of the Bekenstein-Hawking entropy, potentially offering novel observational opportunities for testing modified gravity theories within the semiclassical regime.

\section{Growth of perturbations}
\label{SF}

One of the central challenges in modern cosmology is to understand the mechanisms governing the growth of matter perturbations, as these initial inhomogeneities ultimately lead to the formation of the large-scale structures observed today. It is generally acknowledged that such structures arise from the gravitational amplification of small, primordial density perturbations throughout the cosmic evolution. Over time, these perturbations increase in amplitude, eventually reaching a point where they decouple from the overall cosmic expansion and collapse to form gravitationally bound systems (see Refs.~\cite{Mukhanov:1990me,Malik:2008im} for recent reviews).

A widely used framework for analyzing the evolution of matter perturbations and the process of structure formation within the linear regime is the spherical Top-Hat formalism~\cite{Abramo:2007iu}. This method examines the dynamics of a spherically symmetric, spatially uniform overdensity embedded in an expanding cosmological background, allowing the growth of perturbations within a spherical region to be described through the application of the Friedmann equations~\cite{Planelles:2014zaa}. 
In what follows, we apply this model to explore the influence of the generalized mass-to-horizon entropic model
on the evolution of matter perturbations.

As a first step, let us rewrite the 
modified Friedmann equations \eqref{FM1}-\eqref{FM2} in the equivalent form
\begin{eqnarray}
\label{NF1}
    H^2&=&\left[\frac{4\pi\left(3-n\right)\rho_m}{3\gamma n}
    \right]^{\frac{2}{3-n}}\,,
    \\[2mm]
    \dot H&=&-\frac{4\pi}{\gamma n }\frac{\rho_m}{H^{1-n}}\,,
    \label{NF2}
\end{eqnarray}
which imply
\begin{equation}
    \frac{\ddot{a}}{a} = \dot{H} + H^2=-\left(\frac{3-n}{n}\right)^{\frac{n-1}{3-n}}\left(\frac{4\pi\rho_m}{3\gamma}\right)^{\frac{2}{3-n}},
    \label{ddota}
\end{equation}
where we have implicitly neglected both radiation and the cosmological constant, and considered dust matter. This approximation holds with good accuracy in the redshift interval \( 1 \lesssim z  \lesssim 3\times 10^3 \), during which the expansion is governed predominantly by the matter component.

To make the analysis physically relevant to the growth of cosmic structures, we restrict our attention to the redshift interval \( z \gtrsim 15 \). In this regime, on sufficiently large scales,  the earliest stages of structure formation can be traced through the linear growth of initial overdensities that will eventually collapse into gravitationally bound objects, such as mini-halos of mass \( 10^5 - 10^6\, M_\odot \), which give rise to the first stars (Population~III) at \( z \gtrsim 20 \), and the earliest protogalaxies around \( z \sim 15\text{--}20 \) \cite{White:1977jf, Barkana:2000fd}. At lower redshifts, a growing fraction of perturbations enter the nonlinear regime, where additional effects, such as radiative feedback from reionization and baryonic physics, become important. A proper treatment of these late-time processes requires going beyond the linear approximation and will be addressed in future studies.

To implement the Top-Hat model, we consider a spherically symmetric region with radius $a_p$ and uniform density $\rho_m^{(c)}$. At a given time, this density is taken to be $\rho_m^{(c)} = \rho_m + \delta\rho_m$, where $\delta\rho_m$ denotes the density contrast relative to the background. Within this spherical domain, the matter conservation equation takes the form $\dot \rho_m^{(c)} + 3h\rho^{(c)}_m = 0$ \cite{Abramo:2007iu}, where \( h = {\dot{a}_p}/{a_p} \) is the local expansion rate of the spherical perturbed region with radius \( a_p \). Based on this setup, the second Friedmann equation, when applied to this region, is given by Eq. \eqref{ddota}, with $a\to a_p$ and $\rho_m\to\rho_m^{(c)}$. 

It is convenient at this point to define the density contrast of the fluid as 
\begin{equation}
\label{dm}
    \delta_m=\frac{\rho_m^{(c)}}{\rho_m}-1=\frac{\delta\rho_m}{\rho_m}\,.
\end{equation}
Since the density fluctuation \( \delta\rho_m \) is typically much smaller than the background density \( \rho_m \) in the regime described above, it is reasonable to assume that \( \delta_m \ll 1 \), corresponding to the linear approximation \cite{Abramo:2007iu}. 

Differentiating Eq.~\eqref{dm} with respect to $t$, we find
\begin{equation}
\dot\delta_m=3\left(1+\delta_m\right)\left(H-h\right),
\end{equation}
where we have used the continuity equation. 
Further differentiation gives
\begin{equation}
    \label{dyn}
\ddot\delta_{m}=3\left(1+\delta_m\right)\left(\dot H-\dot h\right)+\frac{\dot\delta^2_m}{1+\delta_m}\,.
\end{equation}
This equation describes the dynamical evolution of matter perturbations within the framework of the spherical Top-Hat model. 

We now observe that Eq. \eqref{dyn} can be further manipulated by using Eq. \eqref{ddota}, yielding
\begin{equation}
\label{diffHh}
 \dot H-\dot h =   h^2-H^2
-\left(\frac{3-n}{n}\right)^{\frac{n-1}{3-n}}\left(\frac{4\pi\rho_m}{3\gamma}\right)^{\frac{2}{3-n}} 
 \left[1-\left(1+\delta_m\right)^{\frac{2}{3-n}}\right].
\end{equation}
In light of the above, this can be expanded to linear order in the density contrast $\delta_m$, obtaining
\begin{equation}
    \dot H-\dot h=h^2-H^2+
    2n^{\frac{1-n}{3-n}}\left(3-n\right)^{\frac{2\left(n-2\right)}{3-n}}
    \left(\frac{4\pi\rho_m}{3\gamma}\right)^{\frac{2}{3-n}}\delta_m+\mathcal{O}(\delta_m^2)\,.
\end{equation}
Therefore, substitution into Eq. \eqref{dyn} gives
\begin{equation}
\label{dynbis}
    \ddot\delta_m+2H\dot\delta_m-c_{\gamma,n}\hspace{0.5mm}\rho_m^{\frac{2}{3-n}}\hspace{0.5mm}\delta_m=0\,,
\end{equation}
where we have introduced the shorthand notation $c_{\gamma,n}=\dfrac{2^{\frac{7-n}{3-n}}\pi^{\frac{2}{3-n}}\left(3-n\right)^{\frac{2\left(2-n\right)}{n-3}}}
{\gamma\left(3\gamma n\right)^{\frac{1-n}{n-3}}}$. It is straightforward to verify that $c_{1,1} = 4\pi$, thereby correctly recovering the standard dynamics in the limit where the generalized entropy \eqref{GMHE} reduces to the Bekenstein–Hawking expression.

For the sake of comparison with the existing literature, it proves convenient to change the time variable to the scale factor $a$. In doing so, we obtain
\begin{equation}
\label{conver}
    \dot \delta_m= a H \frac{d\delta_m}{da}\,, \qquad \ddot \delta_m= a^2H^2\frac{d^2\delta_m}{da^2}+aH\left(H+a\frac{dH}{da}\right)\frac{d\delta_m}{da}\,.
\end{equation}
Substituting into Eq.~\eqref{dynbis} and performing some algebra, we arrive at
\begin{equation}
\label{diffa}
    \frac{d^2\delta_m}{da^2}+\frac{3\left(2-n\right)}{3-n}\frac{1}{a}\frac{d\delta_m}{da}-\frac{6n}{\left(3-n\right)^2}\frac{\delta_m}{a^2}=0\,,
\end{equation}
where Eqs.~\eqref{NF1} and \eqref{NF2} have been explicitly used.

Some comments are in order here: first, we notice that the multiplicative parameter \( \gamma \) does not appear explicitly in the evolution equation of the density contrast, which, on the other hand, is significantly affected by the entropic exponent \( n \).
Once again, it can be observed that, in the limit where \( n \)  assumes unitary value, the result obtained is consistent with the standard evolutionary dynamics, which is described by the well-known equation \cite{Abramo:2007iu}
\begin{equation}
    n=1\,\,\,\,\,\Longrightarrow\,\,\,\,\,  \frac{d^2\delta_m}{da^2}+\frac{3}{2a}  \frac{d\delta_m}{da}-\frac{3}{2a^2}  \delta_m=0\,.
\end{equation}

In order to examine in more detail the effects of the parameter \( n \), let us explicitly solve equation~\eqref{diffa}. As a function of redshift, we obtain $\delta_m(z)=c_1\left(1+z\right)^{\frac{2n}{n-3}}+c_2\left(1+z\right)^{\frac{3}{3-n}}$, 
where $c_i$ $(i=1,2)$ are the integration constants. To avoid unphysically large values of \( \delta_m \) at late times and to recover the standard profile \( \delta_m(z) \sim \frac{1}{1+z} \) in the limit \( n = 1 \), we neglect the decaying mode of the matter density contrast, which corresponds to setting \( c_2 = 0 \) (recall that we are considering only small deviations from the \(\Lambda\)CDM model, i.e., \( |n - 1| \ll 1 \)). Equation \eqref{diffa} then has the solution
\begin{equation}
    \delta_m(z)=c_1(1+z)^{\frac{2n}{n-3}}\,.
\end{equation}
On the other hand, as for the normalization constant \( c_1 \), several approaches have been proposed in the literature. A commonly adopted method involves requiring the matter overdensity to reach a characteristic nonlinear collapse threshold at a given redshift, as typically implemented in studies of structure formation and halo mass functions within the Press–Schechter framework~\cite{Press:1973iz}. While this prescription effectively links the linear growth of perturbations to the onset of gravitational collapse, it inherently relies on extrapolating linear theory beyond its domain of validity.

In contrast, since our analysis remains strictly confined to the linear regime and focuses on the pre-collapse evolution of perturbations, we adopt a physically well-motivated alternative: we fix the normalization using adiabatic initial conditions, \( \delta_m(z_i) = \delta_i \), imposed at a sufficiently high reference redshift \( z_i \), where linear theory is fully applicable \cite{Sheykhi:2022gzb}. This yields $c_1 = \delta_i \left(1 + z_i\right)^{\frac{2n}{3 - n}}$.
Specifically, we choose \( z_i = 1000 \), corresponding to a time shortly after matter–radiation equality and well before recombination, and set the initial amplitude to \( \delta_i \sim 10^{-5} \), in agreement with the magnitude of primordial fluctuations predicted by inflation and supported by measurements of the CMB anisotropies from the Planck mission~\cite{Planck:2018vyg}.

\begin{figure}[t]
\begin{center}
\includegraphics[width=9cm]{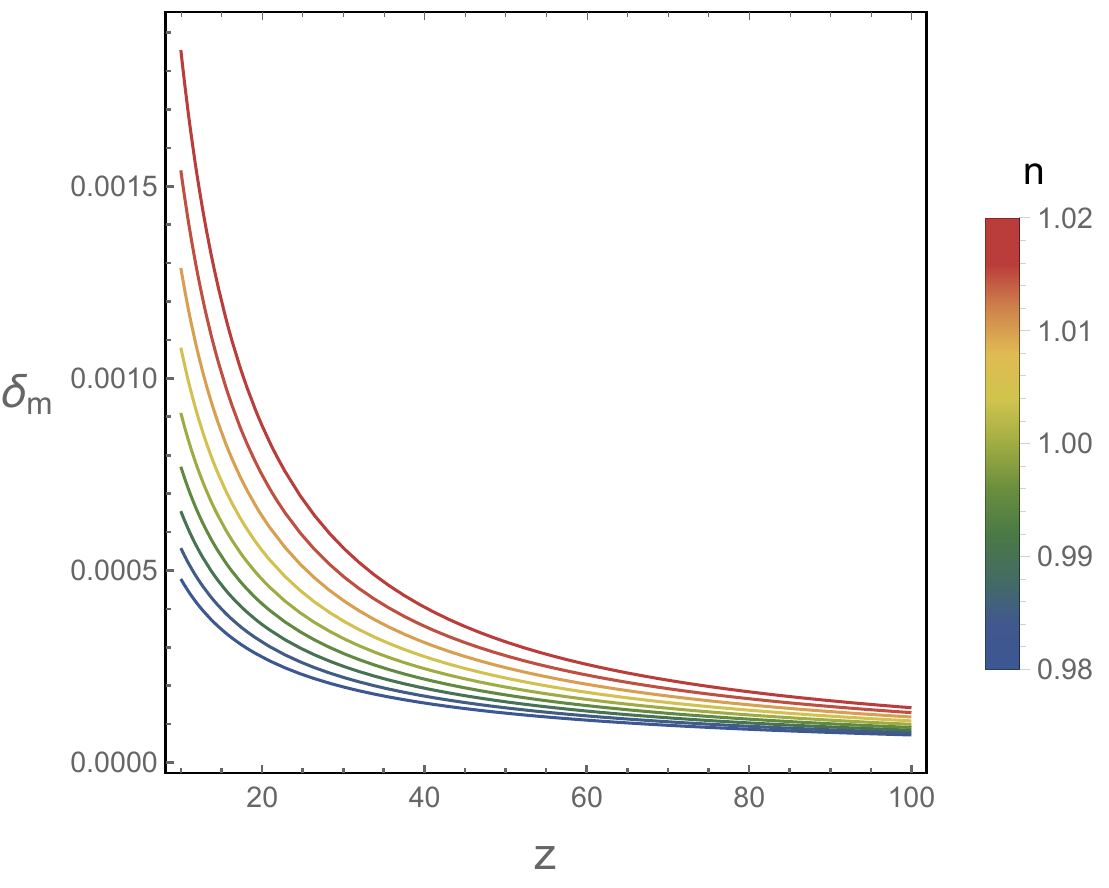}
\caption{Plot of the matter density contrast $\delta_m$ versus $z$, for different values of $n$. }
\label{Pert}
\end{center}
\end{figure}

In Fig.~\ref{Pert}, we plot the matter density contrast \( \delta_m \) as a function of redshift for different values of the entropic exponent \( n \). For \( n > 1 \), the growth of perturbations is faster than in the standard case \( (n = 1) \), particularly at lower redshifts. Conversely, for \( n < 1 \), the growth is suppressed relative to the same benchmark. This behavior can be interpreted in light of the modified cosmological dynamics discussed in Sec.~\ref{ModCos}. Specifically, by looking at Eq. \eqref{dynbis}, we observe that the evolution of \( \delta_m \) is governed by the interplay between two key terms in the perturbation equation: the Hubble friction term, proportional to \( H \dot{\delta}_m \), and the gravitational source term, which scales as \( \rho_m^{2/(3 - n)} \delta_m \). For \( n > 1 \), the expansion rate \( H(z) \) is reduced relative to the standard case (see Fig. \ref{Fig0}), thereby decreasing the damping effect of Hubble friction.  
At the same time, the effective gravitational source term remains consistently greater than its counterpart in the unmodified scenario ($n=1$), 
enhancing the gravitational pull responsible for the growth of overdensities. The combined effect of reduced friction and stronger gravitational driving results in a faster growth of \( \delta_m \). Conversely, for \( n < 1 \), the background expansion is faster and the gravitational amplification weaker, both of which act to suppress the growth of perturbations. 

Therefore, we conclude that extended models of this type may leave distinctive imprints on the early stages of structure formation, potentially affecting the timing, scale dependence and amplitude of the initial collapse of matter overdensities.

\section{Conclusions and Outlook}
\label{Conc}

This work presents an alternative cosmological model inspired by the conjectured connection between thermodynamics and gravitational dynamics. It is known that applying the first law of thermodynamics at the cosmic horizon can lead to the derivation of the Friedmann equations. This framework typically relies on the Bekenstein-Hawking entropy and the Hawking temperature. However, when the entropy expression is generalized beyond its standard form, the same approach yields modified Friedmann equations, giving rise to novel cosmological scenarios. Ensuring the internal consistency of such generalizations requires that modifications to the entropy be accompanied by corresponding changes in other thermodynamic quantities. A recent development in this direction involves modifying the relation between the mass and horizon radius, which leads to a generalized entropy expression characterized by two free parameters, $\gamma$ and $n$~\cite{Gohar:2023hnb,Gohar:2023lta}. The resulting Friedmann equations for a  flat FLRW background were derived in~\cite{Basilakos:2025wwu}, and their predictions were tested against observational data from SNIa, CC and BAO. In the specific case where $\gamma = n = 1$, the standard Bekenstein-Hawking entropy is recovered and the model reduces to the conventional $\Lambda$CDM paradigm. In contrast, for general values of the parameters, the framework leads to richer and non-trivial cosmological dynamics.

In the present analysis, we further investigated the implications of the model with respect to the relic abundance of PGWs and the evolution of the density contrast profile of matter perturbations. In the former context, we set $\gamma=1$ and identified the range in the parameter space of $n$ where deviations from standard cosmology may emerge due to an amplified PGW spectrum. This analysis allowed us to exclude the region $n \lesssim 0.9$ based on current Pulsar Timing Array observations. Although the investigation was carried out by incorporating all relevant corrections at the background level, the resulting bound is non-trivial and may offer a promising framework for testing and constraining the generalized mass-to-horizon entropy relation. Regarding the study of matter perturbation growth and structure formation, we employed the spherical Top-Hat collapse model in the linear regime. We find that the evolution of the matter density contrast \( \delta_m \) is significantly influenced by the value of \( n \), while remaining insensitive to the parameter \( \gamma \).
In particular, values of $n > 1$ ($n < 1$) lead to an enhanced (suppressed) growth rate of $\delta_m$ compared to the standard $\Lambda$CDM scenario. Both of these results can be traced back to modifications in the Hubble expansion rate induced by the generalized mass-to-horizon entropy. Depending on the sign of $n - 1$, this leads to either an enhancement or a suppression of the Hubble rate relative to the $\Lambda$CDM prediction, thereby influencing the dynamics of the early Universe and the growth history of cosmic structures.

Several aspects still require further investigation before the model can be considered a viable description of Nature. For instance, a full dynamical system analysis should be carried out in order to uncover the global features of the cosmic evolution, independently of specific background solutions. 
{Furthermore, our focus was on the pre-collapse regime, where perturbations remain small and the linear approximation is well justified. At lower redshifts, however, an increasing fraction of perturbations evolve into the non-linear regime, where baryonic physics, radiative feedback and mode-coupling effects become non-negligible. Extending our framework beyond linear theory is indeed a natural next step. The entropic modifications we propose alter the background expansion rate and the effective gravitational source term, both of which play a crucial role in non-linear structure formation. As a consequence, one expects that the enhanced (suppressed) linear growth observed for $n>1$ ($n<1$) would translate into an earlier (later) onset of non-linear collapse, thereby modifying halo abundances and clustering statistics. This could directly affect late-time observables such as the matter power spectrum and weak lensing, which are central to the current $\sigma_8$ tension~\cite{DiValentino:2021izs}. In particular, since the $\sigma_8$ (or more precisely $S_8$) tension points to a weaker growth of large-scale structures than predicted by $\Lambda$CDM calibrated on CMB data, our framework suggests that small deviations with $n<1$ could help alleviate the discrepancy by lowering $f\sigma_8(z)$ and thus the inferred value of $S_8$ from large-scale structure observations (see also~\cite{Basilakos:2023kvk}, where the tension was addressed in the context of Tsallis Cosmology). From a phenomenological point of view, this implies that the same entropic modification responsible for changes in the background dynamics may also leave imprints in the clustering sector. A full quantitative assessment requires going beyond the linear Top-Hat approximation (for instance, by incorporating spherical collapse in the non-linear regime to compute the critical overdensity $\delta_c(n,z)$ and virial overdensity $\Delta_{\mathrm{vir}}(n,z)$, and then using Press--Schechter \cite{Press:1973iz} or Sheth--Tormen \cite{Sheth:1999mn} formalisms to predict halo mass functions). This would directly connect the entropic index $n$ to late-time clustering observables such as weak lensing, redshift-space distortions, and cluster counts. 
}

As a further perspective, and with the aim of applying the model consistently across all cosmological epochs, one could consider generalizing the entropy expression in Eq.~\eqref{GMHE} by promoting its parameters to running quantities. In such a framework, $n$ and $\gamma$ would take values close to unity in regimes where the $\Lambda$CDM paradigm is well supported by observational evidence, while allowing for significant deviations only at very low redshift (to address current tensions) and at high redshift (to account for early-universe phenomena in the inflationary era or possible quantum gravitational effects). 
From this perspective, and given the formal similarities between the generalized mass-to-horizon entropy \eqref{GMHE} and the Tsallis entropy discussed in Sec.~\ref{PGW}, valuable hints may be obtained from~\cite{Nojiri:2019skr}, which proposes a Tsallis-like cosmological model with a dynamical entropic index. Work in these directions is currently in progress and will be presented elsewhere.

\acknowledgments 
The research of GGL is supported by the postdoctoral fellowship program of the 
University of Lleida. GGL gratefully acknowledges  the contribution of 
the LISA 
Cosmology Working Group (CosWG), as well as support from the COST Actions 
CA21136 - \textit{Addressing observational tensions in cosmology with 
systematics and fundamental physics (CosmoVerse)} - CA23130, \textit{Bridging 
high and low energies in search of quantum gravity (BridgeQG)} and CA21106 - 
\textit{COSMIC WISPers in the Dark Universe: Theory, astrophysics and 
experiments (CosmicWISPers)}.

\bibliography{Bib}

\end{document}